\documentclass[twocolumn]{article}
\usepackage{amsmath, amssymb}
\usepackage[small]{caption}
\usepackage{graphicx,dblfloatfix}
\usepackage{geometry}
\usepackage{mathtools}
\usepackage{physics}
\usepackage{relsize}
\usepackage{subcaption}
\usepackage{multicol}
\usepackage[inline]{enumitem}
\usepackage{cite}
\usepackage{authblk}

\newcommand*\widebar[1]{%
	\hbox{%
		\vbox{%
			\hrule height 0.5pt 
			\kern0.5ex
			\hbox{%
				\kern-0.2em
				\ensuremath{#1}%
				\kern-0.0em
			}%
		}%
	}%
}

\begin{document}
\onecolumn
\title{\bf Lattice Boltzmann Large Eddy Simulation Model of MHD}
\author[1]{Christopher Flint}
\author[1]{George Vahala}
\affil[1]{Department of Physics, William \& Mary, Williamsburg, Virginia 23185}
\date{\today}
\maketitle

\begin{abstract}

The work of Ansumali \textit{et al.}\cite{Ansumali} is extended to Two Dimensional Magnetohydrodynamic (MHD) turbulence in which energy is cascaded to small spatial scales and thus requires subgrid modeling.  Applying large eddy simulation (LES) modeling of the macroscopic fluid equations results in the need to apply ad-hoc closure schemes.  LES is applied to a suitable mesoscopic lattice Boltzmann representation from which one can recover the MHD equations in the long wavelength, long time scale Chapman-Enskog limit (i.e., the Knudsen limit).  Thus on first performing filter width expansions on the lattice Boltzmann equations followed by the standard small Knudsen expansion on the filtered lattice Boltzmann system results in a closed set of MHD turbulence equations provided we enforce the physical constraint that the subgrid effects first enter the dynamics at the transport time scales.  In particular, a multi-time relaxation collision operator is considered for the density distribution function and a single relaxation collision operator for the vector magnetic distribution function.  The LES does not destroy the property that $\nabla \cdot \mathbf B = 0$ automatically without the need for divergence cleaning.

\end{abstract}
\clearpage
\section{Introduction}\label{intro}

Computational methods are stretched to the limit in trying to solve problems of strong turbulence\cite{Ansumali,ChenDoolen,Carati2001,premnath2009,Ansumali14}. Direct numerical simulations (DNS) attempt to solve the evolution equations by resolving all the scales excited in the turbulence.   Hence, in strong turbulence, DNS will quickly run into resolution problems:  one will not be able to resolve all the excited scales all the way down to the dissipation scales.  Basically, the computational cost of DNS scales as the $Re^3$, where $Re$ is the Reynolds number of the flow.  (The Reynolds number is basically the ratio of the nonlinear to linear terms in the equations).

In Reynolds averaged (RANS) modeling\cite{Pope} one gives up following the time evolution of a particular realization of the turbulence and instead concentrates on the evolution of the mean velocity, i.e., the evolution of the ensemble average flow field.  As is well known, on taking the ensemble average of the momentum equation we are hit with a closure problem:  the evolution equation now involves the unknown higher order Reynolds stress tensor.  The simplest ad-hoc closure scheme is to model these Reynold stresses by a turbulent-viscosity model.

\subsection{Large Eddy Simulations (LES)}
An alternate approach, called the Large Eddy Simulation (LES) model, follows directly the time evolution of the turbulent large scale motions and models the effect of the small scales on the large scales.  It achieves this by introducing a filtering function $ G \left( \vec{r}, \Delta \right) $, which averages out scales of $\order {\Delta}$, giving the filtered velocity

\begin{equation}\label{VelFilter}
\mathbf{\bar u}\left(\vec{r},\Delta\right)=\int_{-\infty}^\infty \mathbf{u} \left(\vec{r}\,'-\vec{r}\,\right)G\left(\vec{r}\,',\Delta\right)d\vec{r}\,'  .
\end{equation}
Again one must resolve a closure problem:  on filtering the Navier-Stokes equations, Eqs. (\ref{NSIncompFiltered}) and (\ref{NSMomentumFiltered}), one obtains the unknown subgrid stress tensor, Eq. (\ref{ReynoldsStress}), $\tau_{\alpha \beta}$.  Thus the filtered equations are (for fluid velocity $\mathbf u$, pressure $p$, viscosity $\nu$)
\begin{eqnarray}
& \label{NSIncompFiltered}
\nabla \cdot \mathbf{\bar u} = 0 \\
& \label{NSMomentumFiltered}
\partial_{t} \, \mathbf{\bar u} + \left( \mathbf{\bar u} \cdot \nabla \right) \mathbf{\bar u}  = -\nabla \bar p + \nu \nabla^2 \mathbf{\bar u} + \nabla \cdot \mathbf{\tau} \\
& \label{ReynoldsStress}
\nabla \cdot \mathbf\tau = \left( \mathbf{\bar u} \cdot \nabla \right) \mathbf{\bar u} - \overline{\left( \mathbf u \cdot \nabla \right) \mathbf u}
\end{eqnarray}
The ad-hoc closure scheme suggested by Smagorinsky\cite{Pope} resolves the subgrid stress tensor Eq. (\ref{SmagorinskyClosure}) by relating it to the (known) filtered strain rate tensor Eq. (\ref{StrainRateTensor}):
\begin{eqnarray}
& \label{SmagorinskyClosure}
\tau_{\alpha \beta} = -2C_S\Delta^2 \left| \, \widebar S \right| \widebar S_{\alpha \beta} = -2\nu_T \, \widebar S_{\alpha \beta} \\
& \label{StrainRateTensor}
\widebar S_{\alpha \beta} = \frac{1}{2} \left( \partial_\beta \bar u_\alpha + \partial_\alpha \bar u_\beta \right) \\
& \left| \, \widebar S \right| = \sqrt{2 \, \widebar S_{\alpha \beta} \, \widebar S_{\alpha \beta}}  \quad   with \quad  \nu_T = C_S\Delta^2 \left| \, \widebar S \right| ,
\end{eqnarray}
where $C_S$ is the unknown Smagorinsky constant, $\nu_T$ the subgrid viscosity, and $\Delta$ the filtering width.

There have been similar attempts to extended Smagorinsky's ideas to MHD\cite{EarlySmagMHD1,EarlySmagMHD2,EarlySmagMHD3,EarlySmagMHD4,Carati2002}. The filtered MHD equations, Eqs. \mbox{(\ref{MHDIncompFiltered}--\ref{MHDMagneticFiltered})}, contain the unknown subgrid stress tensors (\ref{MHDReynoldsFluidStress}, \ref{MHDReynoldsMagneticStress}), $\tau_{\alpha\beta}^{(v)}$ and $\tau_{\alpha\beta}^{(b)}$.
\begin{eqnarray}
& \label{MHDIncompFiltered}
\begin{array}{ccc}
\nabla \cdot \mathbf{\bar u} = 0 &, & \nabla \cdot \mathbf{\overline B} = 0
\end{array} \\
& \label{MHDMomentumFiltered}
\partial_{t} \, \mathbf{\bar u} + \left( \mathbf{\bar u} \cdot \nabla \right) \mathbf{\bar u}  = -\nabla \bar p + \left( \mathbf{\overline B} \cdot \nabla \right) \mathbf{\overline B} + \nu \nabla^2 \mathbf{\bar u} + \nabla \cdot \mathbf\tau^{(v)} \\
& \label{MHDMagneticFiltered}
\partial_{t} \, \mathbf{\overline B} = \left( \mathbf{\overline B} \cdot \nabla \right) \mathbf{\bar u} - \left( \mathbf{\bar u} \cdot \nabla \right) \mathbf{\overline B} + \eta \nabla^2 \mathbf{\bar B} + \nabla \cdot \mathbf\tau^{(b)} ,\\
& \label{MHDReynoldsFluidStress}
\text{where} \quad \nabla \cdot \mathbf\tau^{(v)} = \left[ \left( \mathbf{\bar u} \cdot \nabla \right) \mathbf{\bar u} - \overline{\left( \mathbf u \cdot \nabla \right) \mathbf u} \right] - \left[ \left( \mathbf{\overline B} \cdot \nabla \right) \mathbf{\overline B} - \overline{\left( \mathbf B \cdot \nabla \right) \mathbf B} \right] \\
& \label{MHDReynoldsMagneticStress}
\nabla \cdot \mathbf\tau^{(b)} = \left[ \left( \mathbf{\bar u} \cdot \nabla \right) \mathbf{\overline B} - \overline{\left( \mathbf u \cdot \nabla \right) \mathbf B} \right] - \left[ \left( \mathbf{\overline B} \cdot \nabla \right) \mathbf{\bar u} - \overline{\left( \mathbf B \cdot \nabla \right) \mathbf u} \right]. 
\end{eqnarray}
As a first step one could invoke the Smagorinsky's ad-hoc closure scheme to the filtered MHD equations and so resolve the subgrid stress tensors, Eqs. (\ref{SmagorinskyMHDClosureV}) and (\ref{SmagorinskyMHDClosureB}), by relating them to the mean strain rate tensor, Eq. (\ref{StrainRateTensor}), and the mean current, Eq. (\ref{StrainRateTensorMag}) :
\begin{eqnarray}
& \label{SmagorinskyMHDClosureV}
\tau^{(v)}_{\alpha \beta} = -2C_{Sv}\Delta^2 \left| \, \widebar S \right| \widebar S_{\alpha \beta} = -2\nu_t \, \widebar S_{\alpha \beta} \\
& \label{SmagorinskyMHDClosureB}
\tau^{(b)}_{\alpha \beta} = -2C_{Sb}\Delta^2 \left| \, \widebar j \right| \widebar J_{\alpha \beta} = -2\eta_t \, \widebar J_{\alpha \beta} \\
& \label{StrainRateTensorMag}
\widebar J_{\alpha \beta} = \frac{1}{2} \left( \partial_\beta \widebar B_\alpha - \partial_\alpha \widebar B_\beta \right)
\end{eqnarray}

Another closure scheme proposed by Carati \textit{et. al}\cite{Carati2002} permits the backscatter of energy from the subgrid to resolved scales.  This cross-helicity based closure takes the form
\begin{eqnarray}
& \label{CHClosureV}
\tau^{(v)}_{\alpha \beta} = -2C_{Sv}\Delta^2 \left| \, \widebar S^{\,v}_{\alpha \beta} \, \widebar S^{\,b}_{\alpha \beta} \right| ^{1/2} \widebar S_{\alpha \beta} = -2\nu_t \, \widebar S_{\alpha \beta} \\
& \label{CHClosureB}
\tau^{(b)}_{\alpha \beta} = -2C_{Sb}\Delta^2 \, \mathrm{sgn} \left( \, \widebar j \cdot \widebar \omega \right) \left| \, \widebar j \cdot \widebar \omega \right|^{1/2} \widebar J_{\alpha \beta} = -2\eta_t \, \widebar J_{\alpha \beta} \\
& \begin{array}{cccccc}
\text{where} & \widebar S^{\,v}_{\alpha \beta} = \widebar S_{\alpha \beta} & , & \widebar S^{\,b}_{\alpha \beta} = \frac{1}{2} \left( \partial_\beta \widebar B_\alpha + \partial_\alpha \widebar B_\beta \right) & , & \widebar \omega = \nabla \cross \widebar {\mathbf u}
\end{array}
\end{eqnarray}

\subsection{Lattice Boltzmann MHD}
In typical computational fluid dynamics (CFD) simulations, the Navier-Stokes equations are solved directly.  An alternative approach is an inverse statistical mechanics approach:  move to a lattice kinetic Boltzmann (LB)
representation which under the Chapman-Enskog expansion will recover the desired Navier-Stokes equation.  In particular the difficult-to-resolve convective derivative $(\mathbf{u} \cdot \nabla) \mathbf{u}$ is replaced by a simple advection on the lattice and an algebraic nonlinearity in the LB collision term.  The two basic steps in LB are:
\begin{enumerate*}[itemjoin = \quad , label = (\alph*)]
\item streaming the distribution function to neighboring lattice nodes (i.e., a simple shift operation), and
\item a collisional relaxation operator which requires only local infomation at each spatial node.
\end{enumerate*}
Thus the LB algorithm, in discrete lattice form for a BGK collision operator, is
\begin{equation}
f_{i}(\vec x + \vec c_{i}, t+1) = f_{i}(\vec x, t) - \frac{1}{\tau} (f_{i}(\vec x, t) - f_{i}^{eq}(\rho,\vec u))
\end{equation}
where $\vec c_{i}$ is a lattice streaming velocity, $\tau$ a relaxation time and $f^{eq}$ a suitably chosen relaxation distribution fucntion.   This leads to an extremely parallelizeable and transparent simulation code.  Another important feature of LB is that nonlocal macroscopic derivative terms can be represented by local moments in the LB representation.  For example, it can be shown\cite{ChenDoolen}
\begin{equation}
\sum_{i} c_{i \alpha} c_{i \beta}  f_{i}^{(1)}  =  - \frac {\rho \tau}{3} \left (\frac{\partial u_{\beta}}{\partial x_{\alpha}} + \frac {\partial u_{\alpha}} {\partial x_{\beta}}  \right )
\end{equation}
where $f^{(1)}$ is the perturbed distribution function  $f^{(1)} = f - f^{eq}$.

The extension of Lattice Boltzmann into MHD has been championed by Dellar\cite{DellarMHDLB} wherein the magnetic field is represented by a separate vector distribution function and which is evolved together with the scalar distribution function using the same two basic steps; stream, and collide.   As with the strain rate  being related to moments of the (perturbed) scalar distribution function, it was shown\cite{DellarMHDLB} that $\nabla \cdot \mathbf B$ could be associated with the trace of an antisymmetric tensor.  Thus this LB-MHD extension can enforce  $\nabla \cdot \mathbf B = 0$ to machine accuracy, without the need for any magnetic field divergence cleaning.


Ansumali \textit{et al.}\cite{Ansumali} realised that the 2 limit processes  in LES for an LB representation of Navier-Stokes turbulence  (the Chapman-Enskog expansion in the Knudsen number, Kn, and an expansion in the filter width, $\Delta$) do not commute.  The typical approach of first performing the Chapman-Enskog limit on LB (to reproduce the fluid equations) and then perform the filtering will lead to the closure problem.  However, Ansumali \textit{et al.}\cite{Ansumali} first perform the filter-width expansion $\Delta$ directly on the LB equations.  This was then followed by the usual Chapman-Enskog expansion to recover the final fluid equations.  By requiring that the effects of the subgrid stresses first enter the evolution equations at the transport level one can get a closed form final expression for the LES equations as well as determining the required scaling of the filter width $\Delta$ in terms of the Knudsen number Kn.  It should be noted that  Ansumali \textit{et al.}\cite{Ansumali} restricted their analysis to 2D Navier-Stokes turbulence in which the energy is inverse cascaded to large scales.   It is also interesting to note that Pope\cite{Pope}, has discussed the expansion of the filtered Navier-Stokes equation in terms of the filter width $\Delta$.  The practical problem is that this would force us to perform filtering in the dissipate range - and thereby placing a very heavy burden on the LES solution to be useful in turbulence simulations, basically turning the LES into a DNS.

Here we extend the ideas of Ansumali \textit{et al.}\cite{Ansumali} to MHD.  For simplicity, we restrict our analysis to 2 dimensional (2D) MHD - since in 2D MHD turbulence energy is cascaded to small scales as in 3D MHD turbulence.  Hence there is a need for subgrid modeling in 2D MHD unlike 2D Navier-Stokes turbulence in which there is an inverse cascade of energy to large scales.  In Sec. \ref{Sec:Filters} we introduce the Gaussian filter and perform expansions in the filter width $\Delta$ to evaluate nonlinear filter averages.  In Sec. \ref{Sec:Moments} we discuss the transformation of the LB-MHD algorithm into the moment basis, permitting a multi-relaxation collision model for the density/velocity distributions while we use a single-relaxation model for the vector magnetic distribution function first introduced by Dellar\cite{DellarMHDLB}.  In Sec. \ref{Sec:LES} the LES-LB-MHD model, we first filter the LB-moment equations and present the details, for brevity, of the 3rd moment, $\widebar M_3$.  An expansion is then made in the usual Knudsen number, Kn, to move from the LB-MHD representation to the macroscopic dissipative equations for MHD.  In order that the subgrid effects first affect the dynamics at the transport time scales one must scale the filter width $\Delta \simeq \order{\mathrm{Kn}^{1/2}}$.


\section{Filters and Filter Widths}\label{Sec:Filters}
Consider a filter function,  $ G \left( \vec{r}, \Delta \right) $, which averages over scales of width $\Delta$, so that the filtered field $ \widebar X $ is given by the convolution integral
\begin{equation}\label{Filter}
\widebar{X}\left(\vec{r}\,',\Delta\right)=\int_{-\infty}^\infty X\left(\vec{r}\,'-\vec{r}\,\right)G\left(\vec{r},\Delta\right)d\vec{r}
\end{equation}
where $ \vec{r} $ and $ \Delta $ defines a location on the lattice and the filter width respectively.  For convenience, we shall use the Gaussian filter function which is sharply peaked about $r = 0$
\begin{equation}\label{Gaussian}
G\left(\vec{r},\Delta\right)=\left(\frac{6}{\pi\Delta^{2}}\right)^{\frac{1}{2}}\exp\left(-\frac{6r^2}{\Delta^{2}}\right)
\end{equation}
with the isotropic properties
\begin{equation}
\begin{array}{ccc}
\int_{-\infty}^{\infty} G \left( \vec{r}, \Delta \right) d \vec{r} = 1 \ ,
&\int_{-\infty}^{\infty} G \left( \vec{r}, \Delta \right) \vec{r} d \vec{r} = 0 \ ,
&\int_{-\infty}^{\infty} G \left( \vec{r}, \Delta \right) r_{\alpha} r_{\beta} d \vec{r} = \frac{\Delta^{2}}{12} \delta_{\alpha\beta}
\end{array}
\end{equation}

Taylor expanding the dynamical field $X\left(\vec{r}\,'-\vec{r}\,\right)$ about $\vec{r} = \vec{r}\,'$ in Eq. (\ref{Filter})    and then  performing the Gaussian weighted polynomial integrals one immediately finds\cite{Pope}
\begin{equation}
\widebar X = X + \frac{\Delta^2}{24} \partial_\beta^2 X + \order{ \Delta^4 } 
\end{equation}
Similarly, it can he shown
\begin{equation}\label{FilterExampleSmall}
\overline{\left( XY \right) } = \widebar{X} \, \widebar{Y} + \frac{\Delta^{2}}{12} \left( \partial_{\beta} \, \widebar{X} \right) \left( \partial_{\beta} \, \widebar{Y} \right) + O\left( \Delta ^{4} \right)
\end{equation}
and 
\begin{equation}\label{FilterExampleLarge}
\overline{\left( \frac{XY}{Z} \right) } = \,  \frac{\widebar{X} \, \widebar{Y}}{\widebar{Z}} + \frac{\Delta^{2}}{12 \, \widebar{Z}} \left[ \left( \partial_{\beta} \, \widebar{X} \right) \left( \partial_{\beta} \, \widebar{Y} \right) - \frac{\left( \partial_{\beta} \, \widebar{Z} \right)}{\widebar{Z}} \left( \widebar{X} \left( \partial_{\beta} \, \widebar{Y} \right) + \, \widebar{Y} \left( \partial_{\beta} \, \widebar{X} \right) - \frac{\widebar{X} \, \widebar{Y} \left( \partial_{\beta} \, \widebar{Z} \right) }{\widebar{Z}} \right) \right] + O\left( \Delta ^{4} \right) 
\end{equation}
for arbitary fields $X$, $Y$, and $Z$.

\section{Moment Basis Representation for LES LB-MHD}\label{Sec:Moments}
\begin{figure}[h]
	\centering
	\includegraphics[width=2.25in, height=2.59in, keepaspectratio=false]{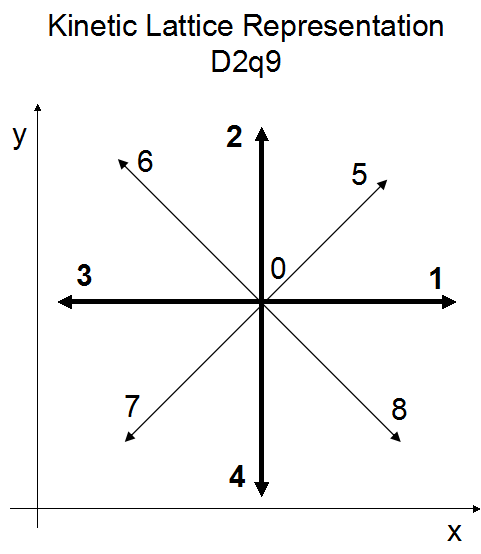}
	\qquad
	\includegraphics[width=2.23in, height=2.54in, keepaspectratio=false]{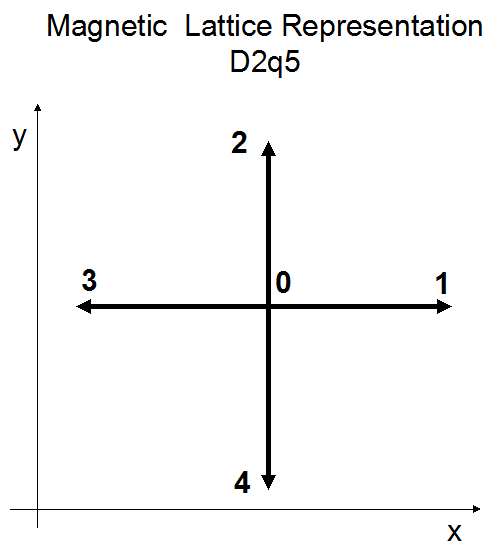}
	\caption[.]{\label{LatticeRep}The lattice vectors for LB-MHD in 2D. 	The kinetic lattice vectors in $D2Q9$ are $\vec c_i = \left(0,0\right),\left(0,\pm 1\right),\left(\pm 1,\pm 1\right) \ , \ i = 0\dots 8$ . The magnetic lattice vectors in $D2Q5$ are $\vec C_k=\left(0,0\right),\left(0,\pm 1\right) \ , \ k = 0\dots 4$ .}
	
\end{figure}

We extend the single relaxation LB-MHD model of Dellar\cite{DellarMHDLB} to incorproate multiple relaxation times (MRT).  We work in 2D for simplicity, and it is readily extended to 3D - but with the complications of a larger number of lattice velocities.  However, unlike the 2D Navier-Stokes work of Ansumali \textit{et. al.}, 2D MHD exhibits the same energy cascade to small scales as in 3D.  
The LB equations for the distribution functions $f_i$, of the density and mean velocity, and $\vec g_k$, for the magnetic field are
\begin{eqnarray}
& \label{LBKinEqn}\left( \partial_t + \partial_\gamma c_{\gamma i} \right) f_i = \sum_j s^{'}_{ij} \left( f_j^{(\mathrm{eq})} - f_j \right)  \\
& \label{LBMagEqn}\left( \partial_t + \partial_\gamma C_{\gamma k} \right) \vec g_k = s^{'}_{m} \left( \vec g_k^{\,(\mathrm{eq})} - \vec g_k \right)
\end{eqnarray}
with the moments $\sum_i f_i = \rho$,  $\sum_i f_i  \vec c_i = \rho \vec u$, and $\sum_k \vec g_k = \vec B$.  In these equations the summation convention is employed on the vector nature of the fields (using Greek indices).  Roman indices correspond to the corresponding lattice vectors for the kinetic velocities $\vec c_i$ and $\vec C_k$ (see Fig. \ref{LatticeRep}).  $s^{'}_{ij}$ and $s^{'}_{m}$ are the collisional relaxation rate tensor for the density and the collisional relaxation rate scalar for the magnetic field distributions, respectively.  The choice of these kinetic relaxation rates will determine the MHD viscosity and resistivity transport coefficients.

To recover the MHD equations,  one must make an appropriate choice of phase space velocity/magnetic field lattice vectors and appropriate relaxation distribution functions.  An appropriate choice for 2D MHD is the 9-bit phase space velocities as seen in \mbox{(Fig. \ref{LatticeRep})} for the density distribution and the simpler 5-bit velocities for the magnetic field distribution.  The simpler lattice for the magnetic field distribution arises since the magnetic field $\vec B$ is the zeroth moment of $\vec g_k$ while the mean fluid velocity is determined from the 1st moment of $f_i$.  To recover the MHD equations in the Chapman-Enskog limit of the (discrete) kinetic equations, an appropriate choice of relaxation distribution functions  $f{_i}^{(eq)}$ and $\vec g_k^{\,(eq)}$ is
\begin{eqnarray}
& D2Q9: \quad f_i^{(eq)} = w_i \rho \left[ 1 + 3\left( \vec c_i \cdot \vec u \right) + \frac{9}{2}\left( \vec c_i \cdot \vec u \right)^2 - \frac{3}{2} \vec u^{\,2} \right] + \frac{9}{2} w_i \left[ \frac{1}{2} \vec B^2 \vec c_i^{\,2} - \left( \vec B \cdot \vec c_i \right)^2 \right]  , i = 0, .. ,8 \\
& D2Q5: \quad \vec g_k^{\,(eq)} = w_k^{'} \!\left[ \vec B + 3 \left\lbrace \left( \vec C_k \cdot \vec u \right) \vec B - \left( \vec C_k \cdot \vec B \right) \vec u \right\rbrace \right] , k = 0, .., 4
\end{eqnarray}


Typically these equations are solved by split-operator methods:  streaming and collisional relaxation.  
In MRT-LB it is more convenient to perform streaming in the distribution space $f_i, \vec g_k$, but to perform the collisional relaxation in moment space.  In particular, it is natural to choose the conservation moments (the zeroth and first moments of the $f_i$ and the zeroth moment of $\vec g_k$)   among the kinetic moments.  The 1-1  constant transformation matrices, $\mathrm T$ and $\mathrm T_{\mathrm m}$, that permit the mapping between the distribution space $(f_i , \vec g_k) $ and the moment space $(M_i , \vec N_k)$ are
 \begin{equation}\label{Transformation}
\begin{array}{ccc}
M_i = \sum_{j=0}^8 \mathrm T_{ij}f_j &, & \vec N_{k} = \sum_{q=0}^4  \mathrm T_{\mathrm m,kq} \vec g_{q}
\end{array}
\end{equation}
with the choice of the moments
\begin{eqnarray}
& \label{FluidTMatrix}
\mathrm T=\left( \begin{array}{c}
\boldsymbol{1} \\ 
c_x \\ 
c_y \\ 
c_xc_y \\ 
c^2_x-c^2_y \\ 
3c_xc^2_y-2c_x \\ 
3c_yc^2_x-2c_y \\ 
4\cdot \boldsymbol{1}-9\left(c^2_x+c^2_y-2c^2_xc^2_y\right) \\ 
4\cdot \boldsymbol{1}-4\left(c^2_x+c^2_y\right)+3c^2_xc^2_y \end{array}
\right)=\left( \begin{array}{ccccccccc}
1 & 1 & 1 & 1 & 1 & 1 & 1 & 1 & 1 \\ 
0 & 1 & 0 & -1 & 0 & 1 & -1 & -1 & 1 \\ 
0 & 0 & 1 & 0 & -1 & 1 & 1 & -1 & -1 \\ 
0 & 0 & 0 & 0 & 0 & 1 & -1 & 1 & -1 \\ 
0 & 1 & -1 & 1 & -1 & 0 & 0 & 0 & 0 \\ 
0 & -2 & 0 & 2 & 0 & 1 & -1 & -1 & 1 \\ 
0 & 0 & -2 & 0 & 2 & 1 & 1 & -1 & -1 \\ 
4 & -5 & -5 & -5 & -5 & 4 & 4 & 4 & 4 \\ 
4 & 0 & 0 & 0 & 0 & -1 & -1 & -1 & -1 \end{array}
\right) \\
\text{and}
& \label{MagTMatrix}
\mathrm T_{\mathrm m}=\left( \begin{array}{c}
\boldsymbol{1} \\ 
C_x \\ 
C_y \\ 
C^2_x \\ 
C^2_y \end{array}
\right)=\left( \begin{array}{ccccc}
1 & 1 & 1 & 1 & 1 \\ 
0 & 1 & 0 & -1 & 0 \\ 
0 & 0 & 1 & 0 & -1 \\ 
0 & 1 & 0 & 1 & 0 \\ 
0 & 0 & 1 & 0 & 1 \end{array}  
\right).
\end{eqnarray}
For 2D LB-MHD the $\mathrm T$-matrix is a $9 \cross 9$ matrix, due to the lattice choice $D2Q9$, and the $\mathrm T_{\mathrm m}$ matrix is a $5 \cross 5$ matrix, due to the lattice choice $D2Q5$ for the magnetic field representation.                
The $x$ and $y$ components of the $9$-dimensional lattice vectors are
\begin{equation}
\begin{array}{ccc}
c_x=\left\{0,1,0,-1,0,1,-1,-1,1\right\} &, &c_y=\{0,0,1,0,-1,1,1,-1,-1\}
\end{array}
\end{equation} 
while the $x$ and $y$ components of the $5$-dimensional lattice vectors for the magnetic distribution are
\begin{equation}
\begin{array}{cccc}
C_x=\left\{0,1,0,-1,0\right\} &, &C_y=\left\{0,0,1,0,-1\right\} &.
\end{array}
\end{equation}                                  

In the moment basis, the collisional relaxation rate tensor in MRT is diagonalized from $s^{'}_{ij}$ to $s_i$ where $\sum_j \sum_k \mathrm{T}_{ij} s^{'}_{jk} f_k = \sum_j \sum_k s_{ij} \delta_{ij} \mathrm{T}_{jk} f_k = s_i \sum_j \mathrm{T}_{ij} f_j$ such that the rank 2 tensor may be reduced to a rank 1 tensor. In the $D2Q9$ phase space, $i=0..8$ for $s_i$ corresponding to the respective moment $M_i$. The collisional relaxation rate scalar for the magnetic field in SRT, $s^{'}_m$, is equal for all magnetic moments, $\vec N_k$, just as it has been for $\vec g_k$, so we will define the relaxation rate for the magnetic field in moment space to be $s_m$ for completeness in notation where $s_m = s^{'}_m$.

The first three fluid moments are nothing but the collisional invariants - being nothing but the conservation of density (the 1\textsuperscript{st} row of the $\mathrm T$-matrix) and the conservation of momentum (the 2\textsuperscript{nd} and 3\textsuperscript{rd} rows of $\mathrm T$). For the $\mathrm T_{\mathrm m}$ matrix only the 1\textsuperscript{st} row is a collisional invariant.
In particular, the moments can be written in terms of the conserved moments:
\begin{equation}\label{KinMomentEq}
\arraycolsep=20pt
\begin{array}{lll}
M_0^{(eq)} = M_0 = \rho & 
M_3^{(eq)} = \frac{\rho u_x \rho u_y}{\rho} - B_x B_y &
M_6^{(eq)} = -\rho u_y \\
M_1^{(eq)} = M_1 = \rho u_x &
M_4^{(eq)} = \frac{\left( \rho u_x \right)^2 - \left( \rho u_y \right)^2}{\rho} - B_x^2 + B_y^2 &
M_7^{(eq)} = -3\frac{\left( \rho u_x \right)^2 + \left( \rho u_y \right)^2}{\rho} \\
M_2^{(eq)} = M_2 = \rho u_y & 
M_5^{(eq)} = -\rho u_x &
M_8^{(eq)} = \frac{5}{3}\rho - 3\frac{\left( \rho u_x \right)^2 + \left( \rho u_y \right)^2}{\rho} \\
\end{array}
\end{equation}
\begin{equation}\label{MagMomentEq}
\arraycolsep=10pt
\begin{array}{llll}
N_{\alpha 0}^{(eq)} = N_{\alpha 0} = B_\alpha & 
N_{\alpha 1}^{(eq)} = \rho u_x B_\alpha - \rho u_\alpha B_x &
N_{\alpha 2}^{(eq)} = \rho u_y B_\alpha - \rho u_\alpha B_y &
N_{\alpha 3}^{(eq)} = N_{\alpha 4}^{(0)} = \frac{B_\alpha}{3} 
\end{array}
\end{equation}

\section{LES at the Kinetic Level}\label{Sec:LES}
\subsection{Filter expansion}
Using the transformations, Eq. (\ref{Transformation}), the LB Eqs. (\ref{LBKinEqn}, \ref{LBMagEqn}) are transformed into the moment basis $M_0, \dots M_8$ and $\vec N_0, \dots \vec N_4$. Thus there will be a set of 9 scalar moment evolution equations for the fluid ($D2Q9$) and 5 vector equations for the magnetic field ($D2Q5$). 
We present the details for just one of these moments, the time evolution of the 3\textsuperscript{rd} fluid moment $M_3$, as the others are done similarly. 
On filtering the evolution equation for $M_3$ 
\begin{equation}\label{FilteredMomentExample}
\partial_t \, \widebar M_3 + \frac{1}{3} \partial_x \left( 2 \, \widebar M_2 + \widebar M_6 \right) + \frac{1}{3} \partial_y \left( 2 \, \widebar M_1 + \widebar M_5 \right) = s_3 \left( \widebar M_3^{(\mathrm{eq})} - \widebar M_3 \right)
\end{equation}
where the problem of closure arises from the evaluation of the $\widebar M_3^{(\mathrm{eq})}$ in the collision term.  From Eq. (\ref{KinMomentEq}) for $\widebar M_3^{(\mathrm{eq})}$, and the filtering expansions Eqs.  (\ref{FilterExampleSmall}, \ref{FilterExampleLarge}), we obtain
\begin{multline}\label{FilteredEq}
\widebar M^{\left(\mathrm{eq}\right)}_3 = \overline{\left( \frac{{\rho u_x} \, {\rho u_y}}{\rho}\right) }-\overline{\left( {B}_x \, {B}_y\right) } \\
= \frac{\widebar{\rho u_x} \, \widebar{\rho u_y}}{\widebar\rho}-{\widebar{B}}_x \, {\widebar{B}}_y + \frac{{\Delta }^2}{12 \, \widebar\rho}\left[\left(\partial_{\beta } \, \widebar{\rho u_x}\right)\left(\partial_{\beta } \, \widebar{\rho u_y}\right)-\frac{\left(\partial_{\beta } \, \widebar\rho\right)}{\widebar\rho}\left(\widebar{\rho u_x}\left(\partial_{\beta } \, \widebar{\rho u_y}\right)+\widebar{\rho u_y}\left(\partial_{\beta } \, \widebar{\rho u_x}\right)-\frac{\widebar{\rho u_x} \, \widebar{\rho u_y}\left(\partial_{\beta } \, \widebar\rho\right)}{\widebar\rho}\right)\right]   \\
  -\frac{{\Delta }^2}{12\ }\left(\partial_{\beta } \, {\widebar{B}}_x\right)\left(\partial_{\beta } \, {\widebar{B}}_y\right) + \order{\Delta^4}.
\end{multline}
It is convenient to rewrite this in the form (for a general moment)
\begin{equation}
\widebar M_i^{(\mathrm{eq})} = M_i^{(\mathrm{eq})} \left( \widebar M_0, \widebar M_1, \widebar M_2, \widebar N_{x0}, \widebar N_{y0} \right) + \Delta^2 \, \widebar M_i^{(\Delta)} 
\end{equation}
where $M_i^{(\mathrm{eq})} \left( \widebar M_0, \widebar M_1, \widebar M_2, \widebar N_{x0}, \widebar N_{y0} \right)$ is just those moment expressions in Eq. (\ref{KinMomentEq}, \ref{MagMomentEq}) but now a function of the filtered conserved moments rather than in their the unfiltered forms, while
 the $\Delta^2 \, \widebar M_i^{(\Delta)}$
 is the term arising from the fact that $\widebar M_i^{(\mathrm{eq})} \neq M_i^{(\mathrm{eq})} \left( \widebar M_0, \widebar M_1, \widebar M_2, \widebar N_{x0}, \widebar N_{y0} \right)$.
 Indeed for the 3rd moment we have
\begin{eqnarray}
& \label{PrefilteredEqEx} M_3^{(\mathrm{eq})} \left( \widebar M_0, \widebar M_1, \widebar M_2, \widebar N_{x0}, \widebar N_{y0} \right) = \dfrac{\widebar{\rho u_x} \, \widebar{\rho u_y}}{\bar\rho}-{\widebar{B}}_x \, {\widebar{B}}_y \\
& \label{DeltaEqEx} \Delta^2 \, \widebar M_3^{(\Delta)} = \dfrac{\Delta^2}{12 \, \bar\rho}\left[\left(\partial_{\beta } \, \widebar{\rho u_x}\right)\left(\partial_{\beta } \, \widebar{\rho u_y}\right)-\dfrac{\left(\partial_{\beta } \, \bar\rho\right)}{\bar\rho}\left(\widebar{\rho u_x}\left(\partial_{\beta } \, \widebar{\rho u_y}\right)+\widebar{\rho u_y}\left(\partial_{\beta } \, \widebar{\rho u_x}\right)-\dfrac{\widebar{\rho u_x} \, \widebar{\rho u_y}\left(\partial_{\beta } \, \bar\rho\right)}{\bar\rho}\right)\right]-\dfrac{{\Delta }^2}{12\ }\left(\partial_{\beta } \, {\widebar{B}}_x\right)\left(\partial_{\beta } \, {\widebar{B}}_y\right)
\end{eqnarray}

\subsection{Knudsen expansion}
 We now expand the filtered LB Eqs. (\ref{FilteredMomentExample}) in the standard way that the fluid equations are derived from the LB by introducing the small parameter $\varepsilon$ which is just the Knudsen number (basically the ratio of the mean free path to the macroscopic length scales). 
 Using multi-time scale analysis, with the advection time scale at $\order {\varepsilon}$ and the transport time scale at  $\order {\varepsilon^2}$, one has
 \begin{equation}\label{ExpandParam}
\begin{array}{ccccccc}
\partial_t \rightarrow \varepsilon \partial_t^{(0)} + \varepsilon^2 \partial_t^{(1)} &, &\partial_\alpha \rightarrow \varepsilon \partial_\alpha &, & \widebar M_i \rightarrow \widebar M_i^{(0)} + \varepsilon \, \widebar M_i^{(1)} +... &, &\widebar {\vec N}_k \rightarrow \widebar {\vec N}_k^{\,(0)} + \varepsilon \, \widebar {\vec N}_k^{\,(1)}  + ...
\end{array}
\end{equation}
In order that the eddy viscosity/resistivity terms come into the filtered fluid equations at the transport time scale and not earlier, one must choose
 $\Delta^2$ to be on the order of the Knudsen number ($\Delta \sim \sqrt{\mathrm{Kn}}$), with $\Delta^2 \, \widebar M_3^{(\Delta)} \sim \varepsilon \, \widebar M_3^{(\Delta)}$. 

The filtered LB equations are now separated into their respective order $\varepsilon$, and $\varepsilon^2$ equations. For $\widebar M_3$ :
\begin{eqnarray}
& \label{LBExpand1} \order {\varepsilon} : \qquad \partial^{(0)}_t \widebar M_3^{(0)} + \frac{1}{3} \partial_x \left( 2 \, \widebar M_2 + \widebar M_6^{(0)} \right) + \frac{1}{3} \partial_y \left( 2 \, \widebar M_1 + \widebar M_5^{(0)} \right) = s_3 \left( \widebar M^{(\Delta)}_3 - \widebar M^{(1)}_3 \right) \\
& \label{LBExpand2} \order{\varepsilon^2} : \qquad \partial^{(0)}_t \widebar M_3^{(1)} + \frac{1}{3} \partial_x \left[ \left(1 - \frac{1}{2} s_6 \right) \widebar M_6^{(1)} \right] + \frac{1}{3} \partial_y \left[ \left(1 - \frac{1}{2} s_5 \right) \widebar M_5^{(0)} \right] + \partial^{(1)}_t \widebar{M^{(0)}_{i}}= - s_3 \left(\widebar M^{(2)}_3 \right)
\end{eqnarray}
where at $\order {1}$, $M_i^{(\mathrm{eq})} \left( \widebar M_0, \widebar M_1, \widebar M_2, \widebar N_{x0}, \widebar N_{y0} \right) = \widebar M_i^{(0)}$.


In general, the unknown terms in the $\order {\varepsilon}$ equations
 must now be determined : $\widebar M_i^{(0)}$, $\widebar M_i^{(\Delta)}$, $\partial_t^{(0)} \widebar M_i^{(0)}$, and $\widebar M_i^{(1)}$.  The $\widebar M_i^{(0)}$ and $\widebar M_i^{(\Delta)}$ terms are determined as above in Eqs. (\ref{PrefilteredEqEx}, \ref{DeltaEqEx}).

The zeroth order time derivatives of the conserved filtered moments $\widebar M_{0..2}$ and $\widebar N_{\alpha 0}$ can be determined by solving the $\order {\varepsilon}$, Eq. (\ref{LBExpand1}), in their corresponding moment representation
\begin{equation}
\partial_t^{(0)} \, \widebar M_0 = - \partial_x \, \widebar M_1 - \partial_y \, \widebar M_2  .
\end{equation}
The remaining zeroth order time derivatives of the non-conserved filtered equilibria $\widebar M_{3..8}$ and $\widebar N_{\alpha\, 1..4}$ can then be found by 
differentiating with respect to the filtered conserved equilibria:
\begin{equation}
\partial_t^{(0)} \widebar M_i^{(0)} \left( \widebar M_0,\, \widebar M_1, \widebar M_2,\, \widebar N_{x0},\, \widebar N_{y0} \right) = \frac{\partial\, \widebar M_i^{(0)}}{\partial\, \widebar M_0} \partial_t^{(0)} \widebar M_0 + \frac{\partial\, \widebar M_i^{(0)}}{\partial\, \widebar M_1} \partial_t^{(0)} \widebar M_1 + \frac{\partial\, \widebar M_i^{(0)}}{\partial\, \widebar M_2} \partial_t^{(0)} \widebar M_2 + \frac{\partial\, \widebar M_i^{(0)}}{\partial\, \widebar N_{x0}} \partial_t^{(0)} \widebar N_{x0} + \frac{\partial\, \widebar M_i^{(0)}}{\partial\, \widebar N_{y0}} \partial_t^{(0)} \widebar N_{y0}
\end{equation}
Since our current LB algorithm itself is accurate to  $\order {\mathrm{Ma}^3}$, where $\mathrm{Ma}$ is the Mach number,  these derivatives need only be evaluated to $\order {\mathrm{Ma}^3}$. Having determined the zeroth order time derivatives of the conserved moments, one substitutes this into the appropriate equation. The solution for $\partial^{(0)}_t \widebar M^{(0)}_3$ is 
\begin{equation}
\partial^{(0)}_t \widebar M^{(0)}_3 = \partial^{(0)}_t \left(\frac{\widebar M_1 \, \widebar M_2}{\widebar M_0} - \widebar B_x \, \widebar B_y + \order{\varepsilon^2} \right) \to 0 + \order{ \mathrm{Ma}^3}.
\end{equation}
Finally, the perturbed moments, $\widebar M_i^{(1)}$, can be calculated by substituting the previous results into the $\order {\varepsilon}$ moment equation (\ref{LBExpand1}) and solving for $\widebar M_i^{(1)}$. The solution for $\widebar M_3^{(1)}$ is
\begin{multline}
\widebar M^{(1)}_3 = - \frac{1}{s_3} \left( \partial^{(0)}_t \widebar M_3^{(0)} + \frac{1}{3} \partial_x \left( 2 \, \widebar M_2 + \widebar M_6^{(0)} \right) + \frac{1}{3} \partial_y \left( 2 \, \widebar M_1 + \widebar M_5^{(0)} \right) \right) + \widebar M_3^{(\Delta)} \\
   = - \frac{1}{3s_3} \left\lbrace \partial_x\widebar M_2 + \partial_y\widebar M_1 \right\rbrace + \frac{\Delta^2}{12 \widebar M_0} \left[ \left( \partial_\beta \widebar M_1 \right) \left( \partial_\beta \widebar M_2 \right) - \frac{ \left( \partial_\beta \widebar M_0 \right) }{\widebar M_0} \left( \widebar M_1 \left( \partial_\beta \widebar M_2 \right) + \widebar M_2 \left( \partial_\beta \widebar M_1 \right) -\frac{\widebar M_1\widebar M_2 \left( \partial_\beta \widebar M_0 \right) }{\widebar M_0} \right) \right] \\
   - \frac{\Delta^2}{12\ } \left( \partial_\beta \, \widebar B_x \right) \left( \partial_\beta \, \widebar B_y \right).
\end{multline}

With the $\order {\varepsilon}$ equations for the conserved moments fully resolved, we must now determine the $\order {\varepsilon^2}$ equations for the conserved moments by solving for the unknown $\partial_t^{(1)} \widebar M_i^{(0)}$. This is determined by substituting these results into the $\order {\varepsilon^2}$ moment equations and then solving for $\partial_t^{(1)} \widebar M_i^{(0)}$. For $\partial_t^{(1)} \widebar M_0$ we find 
\begin{equation}
\left( \partial_t^{(0)} \widebar M_0 + \partial_x \, \widebar M_1 + \partial_y \, \widebar M_2 \right) + \partial_t^{(1)} \widebar M_0 = 0 \quad \to \quad \partial_t^{(1)} \widebar M_0 = 0
\end{equation}

 \subsection{Final filtered LES-MHD equations}\label{Sec:final}
Similarly one proceeds with these steps to determine the filtered MHD equations using the  $\order {\varepsilon}$ and $\order {\varepsilon^2}$ equations for the conserved moments. The final evolution of the continuity $\left( \bar \rho \right)$, momentum $ \left(\, \widebar {\rho \mathbf u} \right) $ and the magnetic field $(\overline{ \mathbf B})$ in our LB-LES-MHD model , after considerable algebra, are (with summation over repeated Greek subscripts)
\begin{equation}\label{Density}
\partial_t \, {\widebar{\rho}} + \nabla \cdot \, \widebar{\rho \mathbf u} = 0  , \qquad   \nabla \cdot \overline{\mathbf B} = 0
\end{equation}
\begin{equation}\label{FinalMomentum}
\begin{split}
\partial_{t} \left(\, \widebar{\rho \mathbf u} \right)+ \nabla \cdot \left( \frac{\widebar{\rho \mathbf u} \ \widebar{\rho \mathbf u}}{\bar\rho} \right) = - \nabla \bar p + \nabla \cdot \left( \overline{\mathbf B} \, \overline{\mathbf B} \right) - \frac{1}{2} \nabla \left( \overline{\mathbf B} \cdot \overline{\mathbf B} \right) + \left( \xi + \frac{1}{3} \nu \right) \nabla \left( \nabla \cdot \widebar{\rho \mathbf u} \right) + \nu \nabla^2 \, \widebar{\rho \mathbf u} \\
   - \nabla \cdot \left\lbrace \frac{6\nu}{6\nu+1} \frac{\Delta^2}{12\bar\rho} \left[ \left( \partial_\beta \left( \, \widebar{\rho \mathbf u} \right) \right) \left( \partial_\beta \left( \, \widebar{\rho \mathbf u} \right) \right) -  \frac{\partial_\beta \bar p}{\bar p} \left( \widebar{\rho \mathbf u} \left( \partial_\beta \left( \, \widebar{\rho \mathbf u} \right) \right) + \left( \partial_\beta \left( \, \widebar{\rho \mathbf u} \right) \right) \widebar{\rho \mathbf u} -  \widebar{\rho \mathbf u} \ \widebar{p \mathbf u} \frac{\partial_\beta \bar p}{\bar p} \right) \right] \right\rbrace \\
- \nabla  \left\lbrace \left( \frac{s_4}{4} + \frac{s_7}{20} - \frac{3 s_8}{10} \right) \frac{\Delta^2}{12\bar\rho} \left[ \left( \partial_\beta \left( \, \widebar{\rho \mathbf u} \right) \right) \cdot \left( \partial_\beta \left( \, \widebar{\rho \mathbf u} \right) \right) - \frac{\partial_\beta \bar p}{\bar p} \left( 2 \, \widebar{\rho \mathbf u} \cdot \left( \partial_\beta \left( \, \widebar{\rho \mathbf u} \right) \right) -  \widebar{\rho \mathbf u} \cdot \widebar{\rho \mathbf u} \frac{\partial_\beta \bar p}{\bar p} \right) \right] \right\rbrace\\
   - \frac{6\nu}{6\nu+1} \frac{\Delta^2}{12} \left\lbrace \frac{1}{2} \nabla \left[ \left( \partial_\beta \overline{\mathbf B} \right) \cdot \left( \partial_\beta \overline{\mathbf B} \right) \right] 
   - \nabla \cdot \left[ \left( \partial_\beta \overline{\mathbf B} \right) \left( \partial_\beta \overline{\mathbf B} \right) \right]
\right\rbrace  ,
\end{split}
\end{equation}
\begin{equation}\label{FinalMagnetic}
\begin{split}
\partial_t \, \overline {\mathbf B} = \nabla \cross \left( \frac{ \widebar{\rho \mathbf u} \cross \overline {\mathbf B}}{\bar\rho} \right) + \eta\nabla^2\,\overline {\mathbf B} + \nabla \cross \left[ \frac{\Delta^2}{12\bar\rho} \frac{6\eta}{6\eta+1} \left\lbrace \left( \partial_\beta \left( \, \widebar{\rho \mathbf u} \right) \right) \cross \left( \partial_\beta \, \overline {\mathbf B} \right) \right. \right. \\
   \left. \left. -  \frac{\partial_\beta \bar p}{\bar p} \left( \left( \, \widebar{\rho \mathbf u} \right) \cross \left( \partial_\beta \, \overline {\mathbf B} \right) + \left( \partial_\beta \left( \, \widebar{\rho \mathbf u} \right) \right) \cross \overline {\mathbf B} -  \frac{ \left( \partial_\beta \bar p \right) }{\bar p} \left( \, \widebar{\rho \mathbf u} \right) \cross \overline {\mathbf B} \right) \right\rbrace \right] .
\end{split}
\end{equation}
In this isothermal model, the equation of state connecting the pressure to the density is $p = \rho  c_s^2 = \frac{\rho}{3}$, in lattice units ($c_s$ is the sound speed).  The transport coefficients (shear viscosity $\nu$, bulk viscosity $\xi$ and resistivity $\eta$) are determined from the LB-MRT relaxation rates:
\begin{equation}
\nu =\frac{1}{3s_3}-\frac{1}{6}=\frac{1}{3s_4}-\frac{1}{6}
\end{equation}
\begin{equation}
\xi =-\frac{1}{9}-\frac{1}{9s_4}-\frac{1}{15s_7}+\frac{2}{5s_8}
\end{equation}
\begin{equation}
\eta =\frac{1}{3s_{m}}-\frac{1}{6}
\end{equation}

If one wished to restrict oneself to a single relaxation (SRT) LB model then  $\left( \frac{s_4}{4} + \frac{s_7}{20} - \frac{3 s_8}{10} \right) = 0$ due to the fact that $s_3 = s_4 = s_5 = s_6 = s_7 = s_8$. This fact leads the $\nabla \left( \widebar {\rho \mathbf u} \cdot \widebar {\rho \mathbf u} \right)$ terms of order $\Delta^2$ to cancel. It should also be noted that one recovers the standard incompressible Lattice Boltzmann model by setting $\xi = \frac{2}{3} \nu$.



\section{Conclusion}
Ansumali \textit{et al.}\cite{Ansumali} have developed a rigorous closure model for 2D Navier-Stokes turbulence by first filtering the LB moment equations and then performing the long-wavelength long-time Knudsen expansion.  The resulting closure model requires that the filter width $\Delta \simeq \order {\mathrm{Kn}^{1/2}}$.  In principle their algorithm can be readily extended to the D3Q27 LB model of 3D Navier-Stokes turbulence where subgrid modeling is now critical.  These authors\cite{Ansumali} then estimate that for 3D Navier-Stokes turbulence, the number of degrees of freedom for the LES kinetic model scales as $Re^{3/2}$ rather than the DNS scaling of $Re^3$ and the total cost of the LES simulation now scales as $Re^{3/2}$ rather than the DNS scaling of $Re^3$.

We have extended the Ansumali \textit{et al.}\cite{Ansumali} LES-LB algorithm to 2D MHD by incorporating the vector distribution function LB representation of Dellar\cite{DellarMHDLB}.  Because there is a direct energy cascade to small scales in 2D MHD, we have here restricted ourselves to 2D turbulence, but extended the LB Navier-Stokes representation to include multiple-collisional-relaxation rates.  The development of a 3D LES-LB-MHD would be somewhat tedious but straightforward.  In our 2D-LES-LB-MHD model, the new subgrid-terms are written in vector form and one notes that they take the form of Smagorinsky tensorial corrections.  This is somewhat to be expected since we have performed Taylor expansions in the filter width.


\section{Acknowledgments}
This was partially supported by grants from the AFOSR and NSF.

\nocite{*}
\bibliographystyle{unsrt}
\bibliography{LESPaperBib}

\end{document}